\documentclass[aps,showpacs, prb]{revtex4-1}
\usepackage{graphicx}
\usepackage{dcolumn}
\usepackage{amsmath}
\usepackage{float}

\makeatother
\begin{document}

\title{Weakly interacting Bose gases below the critical temperature}
\author{E.~Kovalchuk}
\email{KavalchukE@BrandonU.CA}
\author{R.~Kobes}
\email{r.kobes@uwinnipeg.ca}
\affiliation{Physics Department \\and\\ Winnipeg Institute for Theoretical Physics\\
University of Winnipeg\\ Winnipeg, MB\ R3B 2E9\\Canada}

\begin{abstract}
We consider a homogeneous non-ideal Bose gas at nonzero
temperature in equilibrium below the critical temperature $T_C$ in the framework of finite
temperature field theory.
An algorithm is described in which a manageable subset of diagrams can be considered
which, coupled with a self--consistent condition related to the minimization
of the free energy, can be used to obtain physically reasonable results.
\end{abstract}
\maketitle

\section{Introduction}
\label{sec:intro}

Experimental advances on Bose-Einstein condensation (BEC) in trapped
gases of alkali atoms has caused a renewed interest in theoretical investigations
of the phenomenon \cite{alkali-1,alkali-2,alkali-3}.
In particular, the weakly interacting Bose
gas has received much attention, since it is interesting
on its own right and serves as a simple model system \cite{Shi}.

While the basic properties of BECs at zero temperature
are fairly well understood \cite{Pitaevskii}, the theoretical framework 
at finite temperature is less well developed \cite{Andersen}.
Some representative studies in this regard include: a study of the effective potential for a 
weakly interacting Bose gas \cite{haugset}, application of the
background field method to study the interacting Bose gas at finite temperature and 
density \cite{toms}, a study of a two--species homogenous and dilute Bose gas using
the effective potential \cite{pinto}, and an investigation of Bose--Einstein condensation 
of Feshbach molecules at finite temperatures using a mean-field approach \cite{yu} --
a comprehensive review of various studies is also available \cite{jackson}.
In this paper we describe another
approach to the study of
a weakly interacting Bose gas at temperatures below the critical temperature. Based on 
a perturbative diagrammatic expansion of Green functions, the method invokes a self--consistent
condition related to the minimization of the free energy. As will be seen, physically
reasonable results are obtained in regions where a perturbative expansion is expected
to be accurate.

The paper is organized as follows.
In Section \ref{sec:method} we give a general overview of the
procedure and estimate the thermodynamic parameters
of the system under study. The algorithm itself and approximations
made are discussed in Sections \ref{sec:method-procedure} and
\ref{sec:method-discussion}, followed by a discussion of the choice of the
system parameters in Section \ref{sec:method-parameters}.
Section \ref{sec:results} presents our findings and a discussion
of the effects and features observed.
In Section \ref{sec:conclusions} we give some conclusions.

For reasons that will be discussed shortly,
perturbative corrections in this paper are performed within the
thermo-field dynamics (TFD)  framework \cite{Umezawa}. Since this technique
may not be as well known in this context, we discuss it in the Appendix in some detail:
we first review its formal construction in Sections
\ref{app:framework} and \ref{app:greens}, followed by a discussion of the Feynman
diagrams in Section \ref{app:feynman} and a general statement of the
problem of the weakly interacting Bose gas in Section \ref{app:weakly}.
The Appendix finishes with the formulation of the TFD Dyson--Beliaev equations, which serves as
the starting point for considerations in the main body of the paper.

\section{Method}
\label{sec:method}

\subsection{Thermo-field dynamics formalism for a Bose-condensed gas}

In order to incorporate temperature effects,
we will treat the system under study as an open one, exchanging
energy and matter freely with a thermal reservoir, using
the techniques of the field theory at finite
temperature and density \cite{Landsman, kapusta, lebellac}. There are two
main types of finite temperature formalisms: the imaginary--time, or Matsubara,
formalism, and real--time formalisms. The path--integral formulation is a useful
framework for seeing how these two formalisms are related \cite{semenoff, Andersen}. 
The imaginary--time formalism is straightforward,
but an analytic continuation to real time must be done when the calculation of interest
warrants it. The real--time formalism works directly in real time, but at the expense of a doubling
of the number of degrees of freedom, which can make higher--order calculations tedious.
The choice of which formalism to employ is, to a large extent, a matter of personal
preference, as in the end both formalisms must lead to the same physical results. We have found
a real--time formalism, based on an operator--based approach \cite{Pitaevskii, Landau, Abrikosov},
to be convenient for the lower--order calculations done here, as a straightforward interpretation
of the various terms in the perturbative expansion is possible.

The particular formalism we choose is the
thermo-field dynamics (TFD) framework \cite{Umezawa}.
Being a real--time formalism, a doubling of the degrees of freedom is needed at
finite temperature, resulting in a $2\times 2$ matrix propagator.
The diagonal elements of this matrix are Green's functions which describe the
propagation of excited atoms, while the off-diagonal
elements are responsible
for the exchange interactions.
This propagator matrix is subject to the Dyson-Beliaev equations, 
expressed in terms of a $2\times 2$ matrix self-energy \cite{Beliaev}.
These equations are general and could, in principle, be applied to any Bose-condensed
system without reference to whether it is weakly interacting or not
\cite{Shi, Griffin}.
However, except in special cases, a closed--form solution is generally not possible,
and approximation schemes are often invoked. For a perturbative expansion,
choosing an appropriate subset of Feynman diagrams contributing to
the self-energies constitutes a starting point, which we describe
in the next Section.

\subsection{Procedure}
\label{sec:method-procedure}

\begin{figure}[H]
\begin{center}
\includegraphics[width=8cm, angle=0]{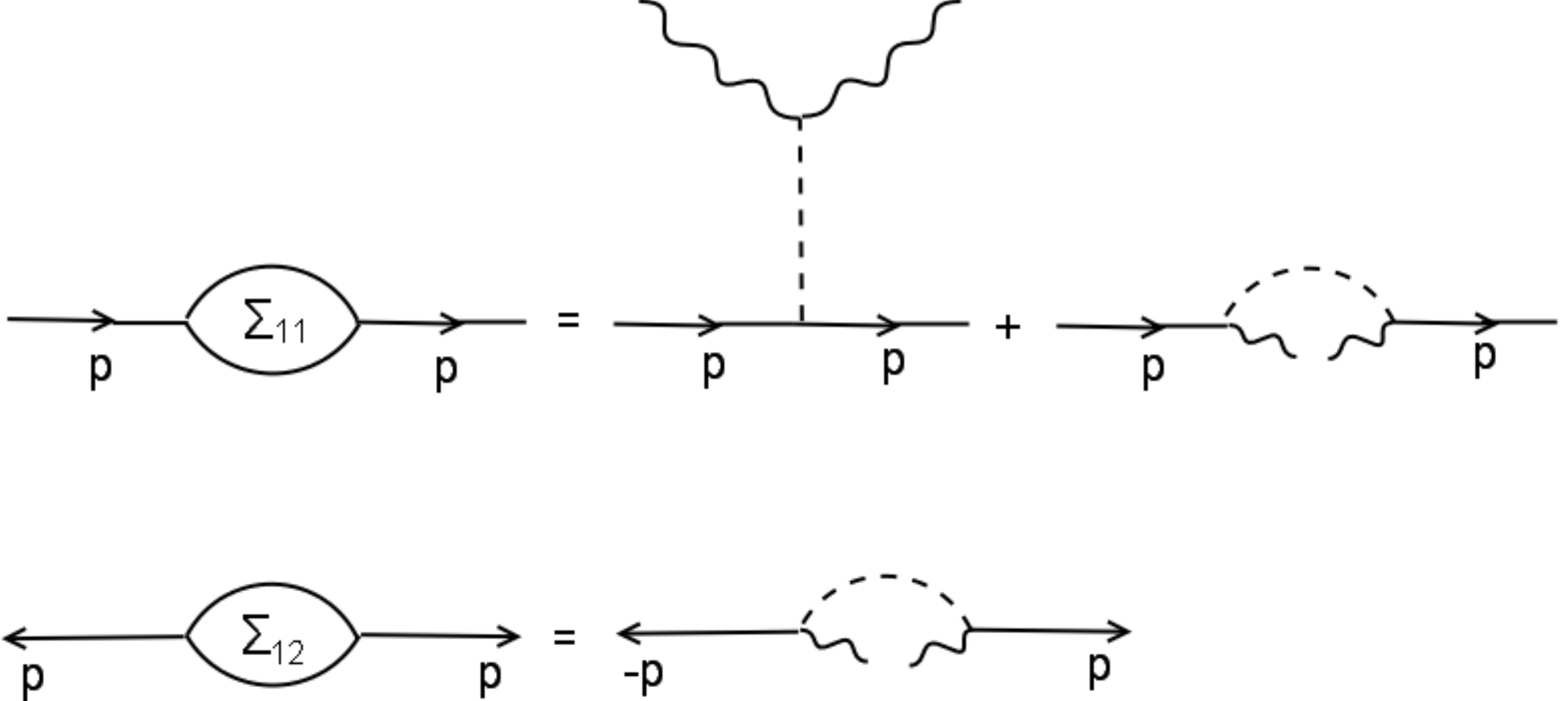}
\end{center}
\caption{First order corrections to the self energy functions of the
         interacting Bose gas in the Bogoliubov approximation. Solid lines
         correspond to propagators of the excited particles, wavy lines to
         the condensed phase ones. Dashed lines represents the interaction potential.
         Details of the notation is found in the Appendix.}
\label{fig:Bogoliubov}
\end{figure}

\begin{figure}[H]
\begin{center}
\includegraphics[width=3cm, angle=90]{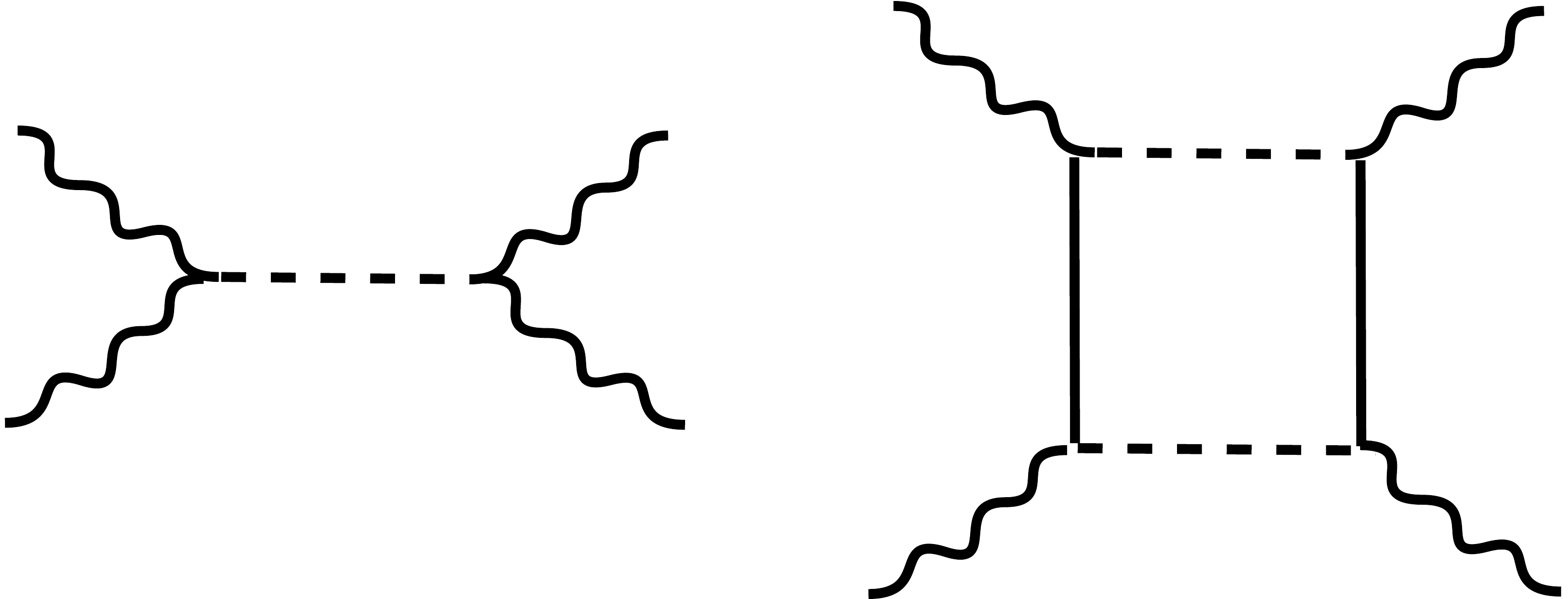}
\end{center}
\caption{First and second order corrections to the free energy of the
         interacting Bose gas in the Bogoliubov approximation.}
\label{fig:free-energy}
\end{figure}

We start with the Dyson-Beliaev equations in TFD for the full propagator matrix,
written in the following form
(see Eqs.~(\ref{a.67}-\ref{a.69}) in the Appendix):
\begin{equation}
 \label{method.1}
 \textbf{G}(p)=\textbf{G}^{(0)} (p)+
 \textbf{G}^{(0)}(p)\cdot
 {\bf \Sigma}(p)\cdot
 \textbf{G}(p),
\end{equation}
where
\begin{eqnarray}
 \label{method.2}
 &&\textbf{G}^{\alpha\beta}(p)=
  \left[
  \begin{array}{ccc}
  G_{11}^{\alpha\beta}(p) \ \
  G_{12}^{\alpha\beta}(p) \\
  G_{21}^{\alpha\beta}(p) \ \
  G_{22}^{\alpha\beta}(p)
  \end{array}
  \right], \\
 \label{method.3}
 &&\textbf{G}^{(0)\alpha\beta}(p)=
  \left[
  \begin{array}{ccc}
  G^{(0)\alpha\beta}(p) \ \
  &&0 \\
  0 \ \
  &&G^{(0)\alpha\beta}(p)
  \end{array}
  \right], \\
 \label{method.4}
 &&{\bf \Sigma}^{\alpha\beta}(p)=
  \left[
  \begin{array}{ccc}
  \Sigma_{11}^{\alpha\beta}(p) \ \
  \Sigma_{12}^{\alpha\beta}(p) \\
  \Sigma_{21}^{\alpha\beta}(p) \ \
  \Sigma_{22}^{\alpha\beta}(p)
  \end{array}
  \right].
\end{eqnarray}
Here $G$ and $G^{(0)}$ are, respectively, the full and free TFD Greens functions
(the latter is given by Eq.~(\ref{a.24})),
while the elements of the $\Sigma$ matrix are the self-energies.
Due to the doubling of degrees of freedom, all of these functions
have a $2\times 2$ matrix structure, indicated by contravariant
indices $\{\alpha, \beta\}$.
Covariant indices $\{1,2\}$ distinguish between the normal and
anomalous types of functions, as described in the Appendix. 
The following identities
help to reduce the number of independent components:
\begin{equation}
 \label{method.5}
 G_{12}^{\alpha\beta}(p)=G_{21}^{\alpha\beta}(-p), ~~~
 G_{22}^{\alpha\beta}(p)=G_{11}^{\alpha\beta}(-p), ~~~
 \Sigma_{12}^{\alpha\beta}(p)=\Sigma_{21}^{\alpha\beta}(-p), ~~~
 \Sigma_{22}^{\alpha\beta}(p)=\Sigma_{11}^{\alpha\beta}(-p).
\end{equation}
Eqs.~(\ref{method.1}) could be solved in principle with respect to the
full propagators (\ref{method.2}), expressed in term of the
self-energy functions (\ref{method.4}). Except in special
cases, however, 
a closed form solution is typically not possible, in which case 
some type of approximation is often used.
In this work, we take into account
only the self-energy functions which correspond to the Bogoliubov
approximation (see Fig.~\ref{fig:Bogoliubov}).
In this approximation one neglects all of the interactions between the excited atoms,
with only those with the condensate taken into account.
With this, we may solve (\ref{method.1}) and find the propagators 
as functions of
two unknown parameters: the condensed phase
density $n_0$ and the chemical potential $\mu$.
In TFD, one may express any physical parameter of interest via the $G^{11}_{11}$
matrix element of the propagator.
In particular, the density $n$ can be written as:
\begin{equation}
\label{method.6}
 n=n_0+\int d\omega ~ d\textbf{p} ~ G_{11}^{11}(\omega,\textbf{p}; n_0, \mu)
\end{equation}
which gives an implicit dependence of $n_0$ on $\mu$.
In order to completely characterize the system we also need, therefore,
an independent way to determine the chemical potential itself.
A common practice here is to use the Hugenholtz-Pines theorem \cite{Hugenholtz},
which relates this parameter to the self-energy functions of the system
\cite{Shi, Andersen}.
However, we proceed in a different
manner and require that the physical
values of the condensate density and the chemical potential
in a state of equilibrium should
minimize the free energy $F$ of the system. In the TFD formalism,
interaction corrections to this potential are given
by a set of vacuum diagrams in which all internal propagators are of
$G^{(0)11}_{11}$ type.
As in the case of the self-energy, we maintain only the minimal possible
subset of diagrams required, which is depicted in Fig.~\ref{fig:free-energy}.
Thus, for example, in the first order of approximation the free energy reads:
\begin{equation}
 \label{method.7}
 F_1=F_0-\frac{1}{2}n_0(\mu)^2 U,
\end{equation}
where
\begin{equation}
 \label{method.8}
 F_0=-\frac{k T}{\lambda^3}g_{5/2}\left(\exp(\mu)\right)
\end{equation}
is the free energy of the ideal Bose gas. Here $k$ is the Boltzmann constant,
$T$ is the temperature, $U$ is the interaction potential,
$g_{5/2}(x)$ is the Bose-Einstein function
\cite{Pathria}, and $\lambda=h/(2\pi m k T)^{1/2}$ is the thermal wavelength,
which, in turn, depends on the mass $m$ of an atom.

In order to keep the procedure consistent, we take into account
the interaction corrections to the self- and free energies
which are of the same expansion order simultaneously.
Our task thus consists of the following steps.
\begin{itemize}
\item  
Given a subset of contributions to the self-energy,
we iteratively solve (\ref{method.6}) to find $n_0$ as a function
of $\mu$, $T$ and $n$.
\item 
This value of $n_0$ is then used to determine the free energy $F$.
\item
The chemical potential is adjusted and this process repeated 
until the minimum of the free energy is found, which thus
corresponds to the physical state of equilibrium.
\end{itemize}
Once the chemical potential and the condensed phase density
as functions of temperature are determined, the rest of the
parameters of interest follow by applying
standard thermodynamic relations.

\subsection{Discussion of the method}
\label{sec:method-discussion}

At first glance, this approach might be seen to suffer from the drawback
that the use the Bogoliubov self-energies 
to explore the thermal properties of the system in a strict perturbative expansion
is only accurate in the vicinity of $T=0$.
However, the determination within our procedure
of the chemical potential
involves a partial resummation of terms in the following sense.
Let $F_n$ be the free energy
in the $n$th--order of approximation,
given by the sum of the zero order free energy $F_0$ and corresponding loop
corrections:
\begin{equation}
 F_n(n_0, \mu)=F_0(n_0, \mu)+F^{(1)}(n_0, \mu)+\ldots +F^{(n)}(n_0, \mu). \nonumber
\end{equation}
The chemical potential is determined by
minimizing the entire right-hand side of this function,
and so places terms of different loop order on the same footing. Thus, a partial
mixing of loop order occurs, but the minimization of the free energy makes the
outcome of this procedure thermodynamically consistent. The subset of the
self--energy diagrams arising in the Bogoliubov approximation is thus regarded
as a minimal starting point in this procedure, and we might anticipate this to potentially provide
reasonable results at temperatures below the critical temperature. Of course, the ultimate
test is to examine the results obtained via this procedure.

\subsection{Model parameters}
\label{sec:method-parameters}

We choose liquid ${^4}$He as a representative system, and so
take the atomic masses and molar volume to be 
$m=6.65\times 10^{-24}$ g and $V=27.6$ cm$^3$/mole respectively
\cite{Pathria}.
Since the momenta involved are sufficiently small due to
the low temperature of the system, we assume
that the Fourier components of the pair-interaction potential $U(\textbf{p})$
could be replaced by the value $U_0$ at $\textbf{p}=0$ \cite{Pitaevskii}. 
Thus, the form of the repulsive interaction potential we use is given by
\begin{equation}
\label{method.9}
 U_0=\frac{4\pi \hbar^2 a}{m},
\end{equation}
where $a$ is the $s$-wave scattering length.
In order to use $U_0$ as a loop expansion parameter, 
the value of $a$ is chosen such that
\begin{equation}
\label{method.10}
 \frac{n U_0}{k T_C}\approx 0.02,
\end{equation}
where $T_C\approx 3.13$ K is the critical temperature of the system.
In other words, the interaction energy between the particles is assumed
to be small compared to the thermal energy.
Furthermore, unless otherwise stated, we will assume that the value of $U_0$  is
unity if it satisfies the assumption of (\ref{method.10}).

\section{Results and discussions}
\label{sec:results}

In this Section we present the results of calculations of the
first and second order interaction corrections
to thermodynamic functions of the ideal Bose gas within this framework.

\begin{figure}[H]
\begin{center}
\includegraphics[width=10cm, angle=0]{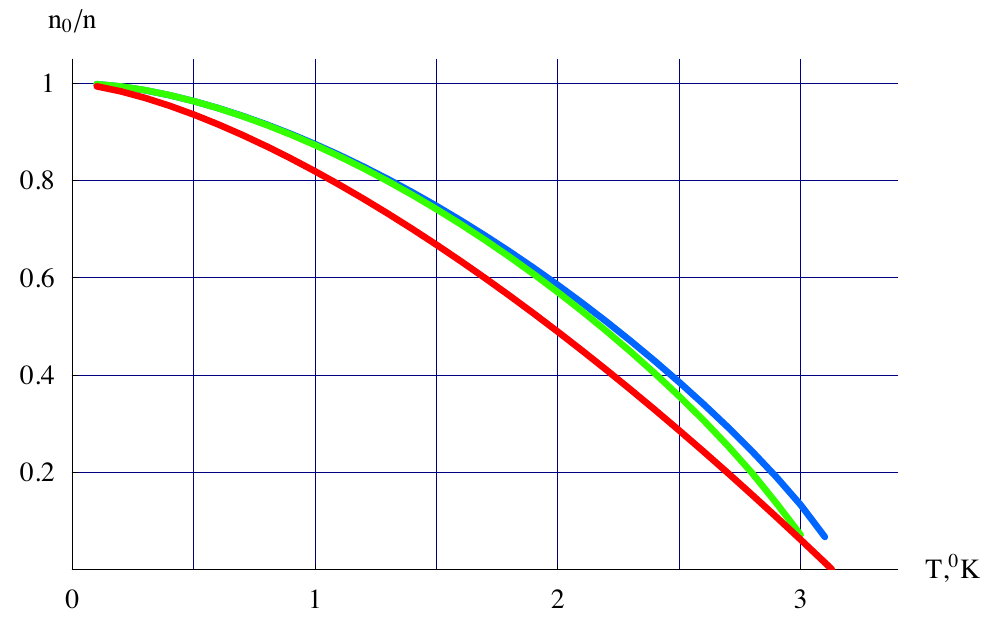}
\end{center}
\caption{The order parameter as a function of temperature.
 Red, blue and green lines correspond, respectively,  to the zeroth,
first and second orders of interaction corrections at unit interaction
 strength.}
\label{fig:n0-phys}
\end{figure}

In Fig.~\ref{fig:n0-phys} we examine
 the dependency of the condensate density on temperature
in the zeroth,
first and second orders of approximation from (\ref{method.6}).
Two features of interest are found here. The first is that the
second order correction does not alter the overall picture too much compared
to the first order one in the whole range of temperatures considered excluding,
possibly, the region close to the critical.
This is what one expects from higher-order loop corrections, and thus provides a
degree of assurance that the perturbative expansion is valid in this region of
temperatures.
The second feature is that it is difficult to
draw any definite conclusions about the system's properties in the critical
region, since the procedure we have followed is not stable there.
This is not a reflection of a particular calculational algorithm,
but reflects the overall drawback of all theories that try to approach the
critical region perturbatively  by taking into account only a limited subset of self-energy
diagrams into account \cite{Andersen}; all such approximations break down near the phase
transition due to infrared divergencies.
This problem might be cured by a selective resummation of diagrams from
all orders of perturbation theory, but since we do not do that here,
we cannot make any definite conclusions about the
critical region, but rather limit our observations in this region
to only general statements.

Thus, we conclude that much of the information concerning
the influence of the weak interaction on the order parameter in the whole
range of temperatures excluding the critical region is contained already
in the first order of approximation. Let us examine this in more detail.

\begin{figure}[H]
\begin{center}
\includegraphics[width=10cm, angle=0]{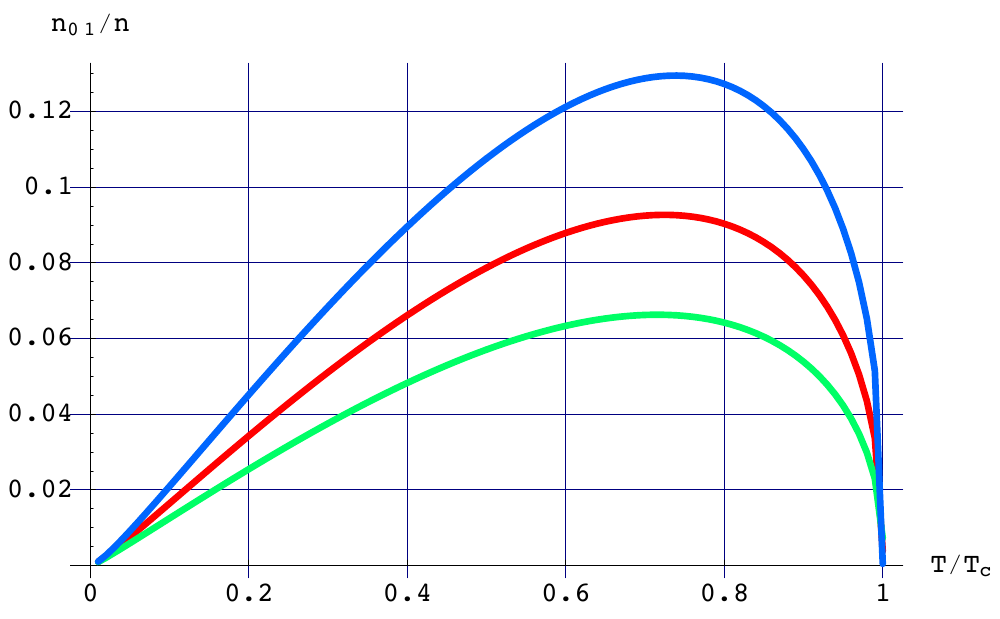}
\end{center}
\caption{The first order interaction correction to the
 condensed phase density as a function of  $T/T_C$.
Green, red and blue lines correspond respectively to the interaction parameter $U_0$
equal to 1/2, 1 and 2.}
\label{fig:n01}
\end{figure}

In Fig.~\ref{fig:n01} the first order interaction correction $n_{0 1}$ from 
(\ref{method.6}) to the
condensed phase density as a function of $T/T_C$ is
represented for three different values of the interaction strength.
First, we notice, that $n_{0 1}$ is positive everywhere except in a
region infinitesimally close to $T=0$.
Thus, at very low temperatures the condensed phase density of the weakly
interacting Bose gas is lower than that of the ideal gas, even at $T=0$.
This effect is a purely entropic nature and appears due to the additional
enthalpic term contributed by the interaction.
This term increases the entropy, which is physically
equivalent to a decrease of the number of condensed particles.

In a strict loop expansion,
the region where $n_{0 1}$ stays negative determines the range of validity
of the Bogoliubov approximation. However, as discussed in
Section \ref{sec:method-discussion}, the partial mixing of loop orders
 makes possible potential conclusions about the system
beyond the scope of the original Bogoliubov construction.
Except in an infinitesimal region near $T=0$, the condensed phase density
of the interacting gas is higher than of its non--interacting counterpart.
An explanation of this could be as follows.
The physics of a non-ideal Bose gases is determined by the
outcome of the competition between the entropy and enthalpy.
While the first favours more particles in a normal phase by increasing the number
of accessible microstates as the temperature increases, the second
tries to put more atoms into the condensate by the use of repulsive interactions.
The point where these two mechanisms balance each other corresponds to the state of
thermodynamic equilibrium. The positivity of
$n_{0 1}$ gives us a clue that the interaction term
gives a greater contribution to the enthalpy than to the entropy in the whole
range of temperatures, starting from the ones just outside of the Bogoliubov
domain up to near $T=T_C$, where the density correction goes to zero.
This seems to imply that the thermal effect is not as significant in an interacting
Bose gas as in an ideal gas.

As is clear from Fig.~\ref{fig:n01}, the bigger the interaction strength, the
bigger is the $n_{0 1}$ correction. Qualitatively, this dependency is nonlinear,
and two-fold. Firstly,
as we observe in Fig.~\ref{fig:n01}, an increase in the magnitude of
the interaction is not accompanied by a proportional increase
of the value of the correction itself. Secondly, the bigger the value of $U_0$, the
greater its change impacts $n_{0 1}$. One may directly observe this effect
from the relative distances between the curves while keeping
 $T/T_C$ fixed.

It is also worth noting that the maximum of $n_{0 1}$ shifts slightly to
higher values of temperatures as $U_0$ increases. This suggests that
the enthalpy is able to dominate the entropy longer the stronger the
interaction is between the atoms.

What is also interesting is the absence of a symmetry between the curves
with respect to the midpoint $T/T_C= 1/2$: the bigger the interaction strength,
the bigger the asymmetry. As discussed above, this is due to the
nonlinear competition between the entropy and enthalpy,
which leads to different slopes on the left and right sides of
the maximum.

\begin{figure}[H]
\begin{center}
\includegraphics[width=10cm, angle=0]{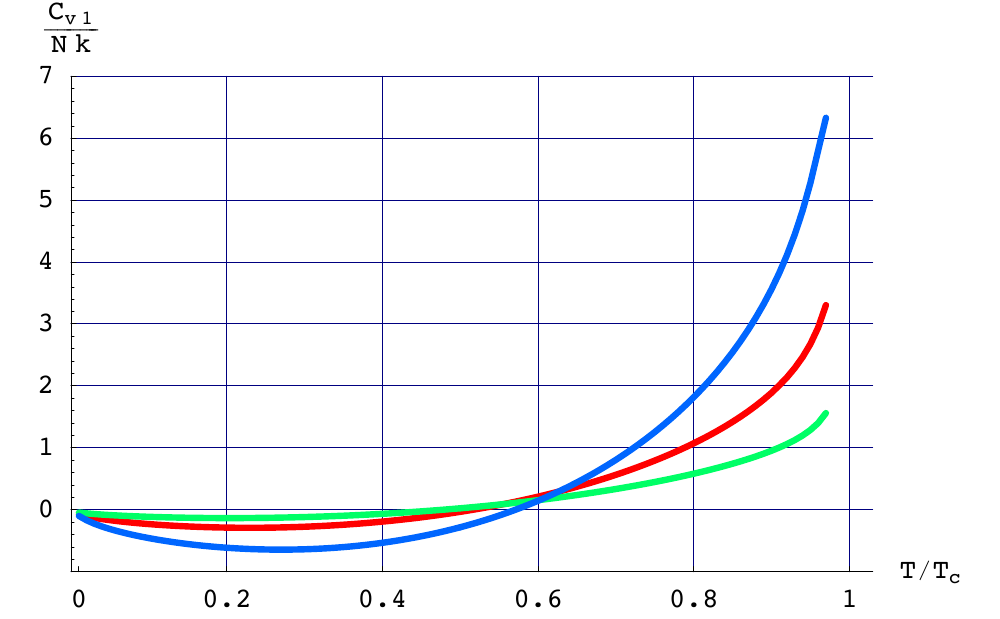}
\end{center}
\caption{The first order interaction correction to the specific heat per particle
 as a function of $T/T_C$.
 Green, red and blue lines correspond respectively
 to the interaction parameter $U_0$
 equal to 1/2, 1 and 2.}
\label{fig:Cv}
\end{figure}

The plot of the dependency of $n_{0 1}$ on $T/T_C$ functionally  and on
$U_0$ parametrically captures most of the physics. We can also
extract some limited information about the behaviour of the system as the temperature
approaches the critical temperature by examining
the first order interaction correction to the specific heat $C_{ V 1}$
per particle
as a function of $T/T_C$. This is shown
in Fig.~\ref{fig:Cv}.
One readily observes the sharp rise of the specific heat as the
temperature nears the critical temperature, with the bigger the interaction parameter,
the sharper the rise. This is the expected qualitative behaviour of
interacting systems near the phase transition point.

\begin{figure}[H]
\begin{center}
\includegraphics[width=10cm, angle=0]{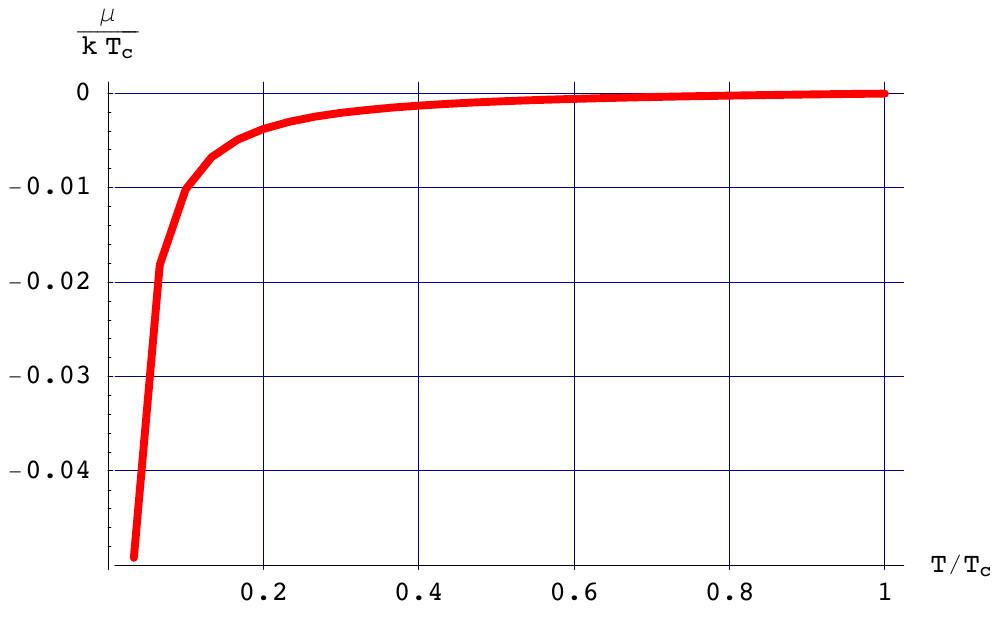}
\end{center}
\caption{The chemical potential in the second order approximation
with unit interaction strength.}
\label{fig:chem-ptl}
\end{figure}

One final feature of this approach should be noted. It turns out that
in the first order of approximation the chemical
potential vanishes. This is in contrast to results obtained within some other
approaches where the chemical potential
of the ideal Bose gas acquires
a small positive correction due to the interaction at first order. In our approach,
the second-order
correction appears significant in this case, as with this,
the chemical potential acquires a small negative value,
comparable in order of magnitude with the interaction strength.
The magnitude is maximal at $T=0$ and quickly drops to zero in
absolute value as the temperature rises, as seen in Fig.~\ref{fig:chem-ptl}.
This feature of second--order effects being significant when first--order effects vanish
is not rare; this is found, for example, in the hard thermal loop resummation of
Braaten and Pisarksi \cite{pisarski} and in certain effective potential calculations
involving electroweak interactions \cite{carrington}.
\section{Conclusions}
\label{sec:conclusions}

Based on a combination of a loop expansion of the self--energy and a
self--consistent minimization of the free energy, we have shown how
thermal properties of the weakly interacting
Bose gas could be explored within a field theoretic framework.
Despite the fact that our procedure uses only the minimal subset of
self-energy and free energy contributions, this allows one to evaluate the
major effects caused by the interaction already in the leading loop 
expansion order.
A notable feature found is an increase in the condensate density,
compared to the non-interacting case, in the whole temperature range
excluding a small region near $T=0$, and a sharp rise of the specific
heat near the phase transition point. These effects clearly demonstrate
that the thermal properties of ideal Bose gases could be changed significantly
by the interaction.

However, as with all such perturbative expansions, the
physically interesting region near the critical temperature remains 
inaccessible by our method.
Summation to all orders of a broad class of self-energy diagrams is likely required 
to make reliable conclusions about  phenomena taking place near the phase
transition point \cite{Andersen}.
\appendix

\section{The TFD framework}

We have chosen the TFD approach as a framework to perform the
calculations in this paper due to the physical interpretation possible in a
perturbative diagrammatic expansion. Although a doubling of the degrees of
freedom is entailed, for lower orders this complication is not extreme, and in any event,
physical quantities should be independent of the formalism used to calculate them.

\subsection{Formal construction of TFD}
\label{app:framework}

The TFD formalism is based on the following arguments \cite{Umezawa}.
Let us imagine an open system, freely exchanging particles and energy
with a heat reservoir, maintained at constant inverse temperature $\beta$.
The process of annihilation of a quanta of energy $\omega(\textbf{k},\beta)$
in such a system can be regarded as the action of an operator $\alpha(\textbf{k})$
which satisfies the Bose commutation relation:
\begin{equation}
 \label{a.1}
 \left[\alpha(\textbf{k}),\alpha^{\dagger}(\textbf{l})\right]=
 \delta(\textbf{k}-\textbf{l}).
\end{equation}
However, unlike the zero temperature case, the operator $\alpha(\textbf{k})$
has an internal structure which incorporates two independent kinds of
thermal effects:
\begin{itemize}
 \item annihilation of a quanta in the system;
 \item creation of the (Dirac) ``holes'' maintained by the reservoir;
\end{itemize}
When the latter process takes place, we say that a $\tilde{\alpha}^{\dagger}$-quantum
(hole) with negative energy $-\hbar\omega(\textbf{k})$ and negative momentum
$-\hbar \textbf{k}$ is created. The former process will be described by
the annihilation operator $\alpha(\textbf{k},\beta)$. Since $\alpha(\textbf{k})$
now performs two independent operations, it can be represented by the
linear combination:
\begin{equation}
 \label{a.2}
 \alpha(\textbf{k})=c(\textbf{k},\beta)\alpha(\textbf{k},\beta)+
 d(\textbf{k},\beta)\tilde{\alpha}^{\dagger}(\textbf{k},\beta),
\end{equation}
where $c(\textbf{k},\beta)$ and $d(\textbf{k},\beta)$ are certain $c$-number
functions. Similarly, the creation operator is
\begin{equation}
 \label{a.3}
 \alpha^{\dagger}(\textbf{k})=c^{*}(\textbf{k},\beta)\alpha^{\dagger}(\textbf{k},\beta)+
 d^{*}(\textbf{k},\beta)\tilde{\alpha}(\textbf{k},\beta),
\end{equation}

By construction, the newly introduced temperature dependent operators satisfy
Bose commutation relations. This suggests that relations (\ref{a.2}-\ref{a.3}) could be
regarded as a canonical transformation, which implies
\begin{equation}
 \label{a.4}
 c(\textbf{k},\beta)c^{*}(\textbf{k},\beta)-d(\textbf{k},\beta)d^{*}(\textbf{k},\beta)=1.
\end{equation}
The phase factors of the above coefficients could be absorbed into those of the
respective operators, so we can choose $c(\textbf{k},\beta)$ and $d(\textbf{k},\beta)$
to be real.

In order to establish their explicit form, let us use the following considerations.
The operator $\alpha^{\dagger}(\textbf{k})$ takes care of the excitations
in the system. Therefore, the quantum number density operator is
\begin{equation}
 \label{a.5}
 N(\textbf{k})=\alpha^{\dagger}(\textbf{k})\alpha(\textbf{k}).
\end{equation}
Since the model we consider is in the grand canonical ensemble,
we may require that the vacuum average of this operator
should be equal to the corresponding value of the number density
in statistical mechanics:
\begin{equation}
 \label{a.6}
 n(\textbf{k})=\frac{(2\pi)^3}{V}\langle 0,\beta|N(\textbf{k})
 |0,\beta\rangle=f_B(\omega),
\end{equation}
with
\begin{equation}
 \label{a.7}
 f_B(\omega)=\frac{1}{e^{\beta\omega(\textbf{k},\beta)}-1}.
\end{equation}
This gives
\begin{equation}
 \label{a.8}
 c(\textbf{k},\beta)=\sqrt{1+f_B(\omega)}, ~~~~
 d(\textbf{k},\beta)=\sqrt{f_B(\omega)}.
\end{equation}
Eq.~(\ref{a.6}) constitutes
the axiom which determines the temperature of the system in a thermal
equilibrium state. It can easily be extended to any operator consisting of
$\alpha$ and $\alpha^{\dagger}$, say $A(\alpha,\alpha^{\dagger})$. Thus, the
vacuum expectation value $\langle 0,\beta|A(\alpha,\alpha^{\dagger})
|0,\beta\rangle$ is equal to the ensemble average of
$A(\alpha,\alpha^{\dagger})$. In this way, the statistical mechanics of a
quantum many-body system becomes a quantum field theory realized in a
temperature-dependent Fock space. This formalism is called {\it thermo--field
dynamics}.

The form of Eqs.~(\ref{a.2}, \ref{a.3}) suggests that tilde and non-tilde operators
are related to each other. Indeed, there is a general TFD
``tilde-conjugation'' operation which is valid not only for the elementary
creation and annihilation operators, but also for arbitrary compound
functions $O_1$ and $O_2$:
\begin{eqnarray}
 \label{a.9}
 &&\widetilde{O_1O_2}=\tilde{O}_1\tilde{O}_2, \\
 \label{a.10}
 &&\widetilde{c_1O_1+c_2O_2}=c^{*}_1\tilde{O}_1+c^{*}_2\tilde{O}_2, \\
 \label{a.11}
 &&\tilde{\tilde{O}}=O.
\end{eqnarray}
Constructing the free TFD Hamiltonian $\hat{H}_0$ is now straightforward:
\begin{equation}
 \label{a.12}
 \hat{H}_0=H_0-\tilde{H}_0.
\end{equation}
with
\begin{equation}
 \label{a.13}
 H_0=\hbar\int d~\textbf{k} ~ \omega(\textbf{k})
 \alpha^{\dagger}(\textbf{k},\beta)\alpha(\textbf{k},\beta), ~~~
 \tilde{H}_0=\hbar\int d~\textbf{k} ~ \omega(\textbf{k})
 \tilde{\alpha}^{\dagger}(\textbf{k},\beta)\tilde{\alpha}(\textbf{k},\beta)
\end{equation}
or, equivalently, with the use of the canonical transformation
\begin{equation}
 \label{a.14}
 H_0=\hbar\int d~\textbf{k} ~ \omega(\textbf{k})
 \alpha^{\dagger}(\textbf{k})\alpha(\textbf{k}), ~~~
 \tilde{H}_0=\hbar\int d~\textbf{k} ~ \omega(\textbf{k})
 \tilde{\alpha}^{\dagger}(\textbf{k})\tilde{\alpha}(\textbf{k}).
\end{equation}

\subsection{Green's functions in TFD}
\label{app:greens}

Let us now consider the simple example of a free field at finite temperature.
Suppose such a field $\phi^0$ satisfies the equation $\lambda(\partial)\phi^0=0$,
with 
\begin{equation}
 \label{a.15}
 \lambda(\partial)=i\frac{\partial}{\partial t}-\omega(\nabla).
\end{equation}
The expressions for the tilde- and non-tilde Hamiltonians read:
\begin{eqnarray}
 \label{a.16}
&& H_0=\int d \textbf{x} \ \phi^{0\dagger}\omega(\nabla)\phi^0,\\
 \label{a.17}
 &&\tilde{H}_0=\int d \textbf{x} \ \tilde{\phi}^{0\dagger}\omega(-\nabla)\tilde{\phi}^0,
\end{eqnarray}
where the relation $\omega^{*}(\nabla)=\omega(-\nabla)$ from the
tilde-conjugation rules (\ref{a.9}-\ref{a.11}) has been used.
The corresponding equations of motion
\begin{equation}
 \label{a.18}
 i\hbar \frac{\partial}{\partial t}\phi^0(\textbf{x})=\left[
  \phi^0(\textbf{x}),H_0
 \right],
\end{equation}
\begin{equation}
 \label{a.19}
 i\hbar\frac{\partial}{\partial t}\tilde{\phi}^0(\textbf{x})=-\left[
 \tilde{\phi}^0(\textbf{x}),\tilde{H}_0
 \right].
\end{equation}
have the formal solution:
\begin{eqnarray}
 \label{a.20}
 &&\phi^0(\textbf{x})=\frac{\sqrt{\hbar}}{(2\pi)^{3/2}}
 \int d \textbf{k} \ \alpha(\textbf{k})\exp[i\textbf{k}\cdot\textbf{x}-i\omega(\textbf{k})t],\\
 \label{a.21}
 &&\tilde{\phi}^0(\textbf{x})=\frac{\sqrt{\hbar}}{(2\pi)^{3/2}}\int d \textbf{k} \
 \tilde{\alpha}(\textbf{k}) \exp[-i\textbf{k}\cdot\textbf{x}+i\omega(\textbf{k})t].
\end{eqnarray}
It is useful to
simplify the notation by introducing the following thermal doublet symbol:
\begin{equation}
 \label{a.22}
 \left(
 \begin{array}{ccc}
   \phi^1  \\ \phi^2
 \end{array}
 \right) =
 \left(
  \begin{array}{ccc}
   \phi^0  \\ \tilde{\phi}^{0\dagger}
  \end{array}
 \right).
\end{equation}
The causal two-point function is then defined by
\begin{eqnarray}
 \label{a.23}
 G^{(0)\alpha\beta}(x-y)&=&
 \langle 0,\beta|T[\phi^{\alpha}(x),
 \phi^{\beta\dagger}(y)]|0,\beta\rangle = \nonumber \\ \
 &=&\theta(t_x-t_y)\langle 0,\beta|\phi^{\alpha}(x)\phi^{\beta\dagger}|0,\beta\rangle
 +\theta(t_y-t_x)\langle 0,\beta|\phi^{\beta\dagger}(y)
 \phi^{\alpha}(x)|0,\beta\rangle
 \label{1.10.46}
\end{eqnarray}
where the symbol $T$ denotes ordering along the TFD time contour in the complex plane
\cite{semenoff}, and the
average is over the temperature-dependent vacuum.
Using (\ref{a.20}, \ref{a.21}) and the canonical
transformations (\ref{a.2}, \ref{a.3}), we find  that in the
momentum representation the above function reads:
\begin{eqnarray}
 \label{a.24}
  G^{(0)\alpha\beta}(k_0, \textbf{k})&=&  \left( 
  \begin{array}{ll}
 {  \displaystyle
  \frac{c^2_B(\omega)}{k_0-\omega+i\epsilon}-\frac{d^2_B(\omega)}{k_0-\omega-i\epsilon} }&
{  \displaystyle
  \frac{c_B(\omega)d_B(\omega)}{k_0-\omega+i\epsilon}-\frac{c_B(\omega)d_B(\omega)}
  {k_0-\omega-i\epsilon} } \\
   & \\
{  \displaystyle
  \frac{c_B(\omega)d_B(\omega)}{k_0-\omega+i\epsilon}-\frac{c_B(\omega)d_B(\omega)}
  {k_0-\omega-i\epsilon} } &
{  \displaystyle
  \frac{d^2_B(\omega)}{k_0-\omega+i\epsilon}-\frac{c^2_B(\omega)}{k_0-\omega-i\epsilon}}
  \end{array}
 \right) 
\end{eqnarray}
where $\omega=\omega(\textbf{k})$ and $c_B(\omega), d_B(\omega)$ are given by (\ref{a.6}).
This free propagator (\ref{a.24}) is the building block for
the development of perturbation theory, which we now discuss.

\subsection{Feynman diagrams in TFD}
\label{app:feynman}

Perturbative calculations in TFD are similar to
those at zero temperature, and indeed, one of the virtues of the formalism is
that the machinery of renormalization and the renormalization group
equations are readily incorporated.
The doubling of degrees of freedom discussed above leads to a $2\times2$
matrix structure of propagators and self energies.
This feature implies that in higher orders a much larger number of
diagrams has to be taken into account, compared to the vacuum theory,
and calculations may become cumbersome. However, for calculations to low
orders, this is not a serious problem.

In general, Feynman diagrams turn out to be topologically and
combinatorically identical to those of the corresponding
zero-temperature quantum field theory, but generalized to a two-component
formalism.
The first (type 1) component represents, by construction, the ``physical'' field,
describing propagation of excitations in the system.
The second (type 2) component corresponds to a
``thermal ghost field'', corresponding to holes, maintained by the reservoir.
In TFD, two types of vertices occur:
one type describing the interactions of the type 1 fields, and the other
of the type 2 fields.
The Feynman rules for these two types of fields differ only by a sign in the
case of even interactions.
There is no direct coupling between the two types of fields, but they can
propagate into each other because of the non-diagonal 1--2 and 2--1 elements of the
propagator matrix.
The 1--1 and 2--2 components correspond to physical and ghost propagators.
Therefore, external lines of {\it physical} Green's function are always of type 1.
In order to find a particular $n$-point Green's function in momentum space,
one must draw all topologically distinct diagrams with $n$ external points
of type 1, and then sum over internal vertices of types 1 and 2.
The ultimate goal is to
compute the 1--1 matrix component of the Green's function, which
can then be used to calculate physical quantities of interest.

\subsection{Weakly interacting Bose gas in TFD}
\label{app:weakly}

Let us now formulate the general problem of
a weakly interacting Bose gas within the TFD framework.

The system we study is a dilute gas
consisting of $N$ atoms obeying
Bose statistics, enclosed in a box of volume $V$ and interacting
through a two-body potential $U(\textbf{x}-\textbf{y})$. The
interaction is assumed to be weak which, together with the
low density of the gas, allows us to neglect higher-order
interactions and treat the system's properties perturbatively.
For simplicity, the atoms are assumed to have zero net spin, and
are described by the boson field operators $\psi(\textbf{x}),
\psi^{\dagger}(\textbf{x})$ subject to the Bose commutation
relations:
\begin{eqnarray}
 \label{a.25}
& & [\psi(\textbf{x}),\psi(\textbf{y})]=0, \\
& &  [\psi^{\dagger}(\textbf{x}),\psi^{\dagger}(\textbf{y})]=0, \\
& & [\psi(\textbf{x}),\psi^{\dagger}(\textbf{y})]=
 \delta(\textbf{x}-\textbf{y}) 
\end{eqnarray}
The system is immersed in a large heat reservoir, with which
it can freely exchange both energy and particles. In a state of
equilibrium both the system and the reservoir possess the same chemical
potential $\mu$ and inverse temperature $\beta$. In TFD 
it is assumed that the reservoir, unlike the system, maintains the
negative-energy holes, represented by the operators obtained
by the tilde-conjugation operation of
(\ref{a.9}-\ref{a.11}). They satisfy the same commutation
relations
as the regular operators, while all commutators between the tilde-
and non-tilde operators vanishing by construction.

In equilibrium, the system contains a mixture of
particles and holes. Therefore, its Hamiltonian $\hat{H}$ is given by
\begin{equation}
 \label{a.27}
 \hat{H}=K-\tilde{K},
\end{equation}
where
\begin{eqnarray}
 \label{a.28}
 K=H_0-\mu N&=&\int d\textbf{x} \ \psi^{\dagger}(\textbf{x})
\left[
 -\frac{\hbar^2}{2m}\nabla^2-\mu
\right]\psi(\textbf{x})\\  \nonumber
&+&\frac{1}{2}\int d\textbf{x} \int d\textbf{y}\psi^{\dagger}(\textbf{x})
\psi^{\dagger}(\textbf{y})U(\textbf{x}-\textbf{y})\psi(\textbf{y})
\psi(\textbf{x})
\end{eqnarray}
is the regular zero-temperature Hamiltonian describing the particles
of the system
and $\tilde{K}$ is its tilde-conjugate, describing the holes:
\begin{eqnarray}
 \label{a.29}
 \tilde{K}=\tilde{H_0}-\mu \tilde{N}&=&\int d\textbf{x} \
 \tilde{\psi}^{\dagger}(\textbf{x})
\left[
 -\frac{\hbar^2}{2m}\nabla^2-\mu
\right]\tilde{\psi}(\textbf{x}) \\ \nonumber
&+&\frac{1}{2}\int d\textbf{x} \int d\textbf{y}\tilde{\psi}^{\dagger}(\textbf{x})
\tilde{\psi}^{\dagger}(\textbf{y})U(\textbf{x}-\textbf{y})
\tilde{\psi}(\textbf{y})\tilde{\psi}(\textbf{x}).
\end{eqnarray}
The chemical potential $\mu$ is chosen such that the vacuum average of the
atomic number operator is equal to the total number of particles in the system:
\begin{equation}
 \label{a.30}
 \langle\hat{N}\rangle=N.
\end{equation}
With the Hamiltonian
$\hat{H}$, one may introduce the time-dependent Heisenberg picture for any
Schr\"odinger operator $O(\textbf{x})$ and its tilde-conjugate
$\tilde{O}(\textbf{x})$:
\begin{eqnarray}
 \label{a.31}
 O(\textbf{x},t)&\equiv& e^{i K t} O(\textbf{x}) e^{-i K t}, \\
 \label{a.32}
 \tilde{O}(\textbf{x},t) &\equiv& e^{-i \tilde{K} t}
 \tilde{O}(\textbf{x}) e^{i \tilde{K} t}.
\end{eqnarray}
In particular, the field operators have a time-dependence given by
\begin{eqnarray}
 \label{a.33}
 \psi(\textbf{x},t) &=& e^{i K t} \psi(\textbf{x}) e^{-i K t}, \\
 \psi^{\dagger}(\textbf{x},t)&=&e^{-i K t} \psi^{\dagger}(\textbf{x}) e^{i K t}; \\
 \label{a.34}
 \tilde{\psi}(\textbf{x},t)&=&e^{-i \tilde{K} t}
 \tilde{\psi}(\textbf{x}) e^{i \tilde{K} t}, \\
 \tilde{\psi}^{\dagger}(\textbf{x},t) &=&
 e^{i \tilde{K} t} \tilde{\psi}^{\dagger}(\textbf{x}) e^{-i \tilde{K} t}.
\end{eqnarray}
In order to simplify the notation, we introduce the
thermal doublet symbol (see (\ref{a.22}))
\begin{equation}
 \label{a.35}
 \left(
 \begin{array}{ccc}
   \psi^1(\textbf{x},t)  \\ \psi^2(\textbf{x},t)
 \end{array}
 \right) =
 \left(
  \begin{array}{ccc}
   \psi(\textbf{x},t)  \\ \tilde{\psi}^{\dagger}(\textbf{x},t)
  \end{array}
 \right).
\end{equation}
The single-particle Green's function is then defined as:
\begin{equation}
 \label{a.36}
 G^{\alpha\beta}(\textbf{x}, t; \textbf{x}', t')=
 \langle 0,\beta|T[\psi^{\alpha}(\textbf{x}, t),
 \psi^{\beta\dagger}(\textbf{x}', t')]|0,\beta\rangle,
\end{equation}

For our purposes, it is more convenient to work in the single-particle
momentum representation. In this representation, the field operators
$\psi(\textbf{x}), \tilde{\psi}(\textbf{x})$ can be written as
\begin{eqnarray}
 \label{a.37}
 &&\psi(\textbf{x})=\frac{1}{(2\pi)^{3/2}}
 \int d\textbf{k} \ a_{\textbf{k}}\phi_{\textbf{k}}(\textbf{x}),\\
 \label{a.38}
 &&\tilde{\psi}(\textbf{x})=\frac{1}{(2\pi)^{3/2}}\int d\textbf{k} \
 \tilde{a}_{\textbf{k}} \tilde{\phi}_{\textbf{k}}(\textbf{x}),
\end{eqnarray}
where $a_{\textbf{k}} (a^{\dagger}_{\textbf{k}})$ and
$\tilde{a}_{\textbf{k}} (\tilde{a}^{\dagger}_{\textbf{k}})$
are the creation (annihilation) operators for a particle and hole in
single-particle states $\phi_{\textbf{k}}(\textbf{x})$ and
$\tilde{\phi}_{\textbf{k}}(\textbf{x})$, respectively.
These obey Bose commutation relations:
\begin{eqnarray}
 \label{a.39}
 &&[a_{\textbf{k}}, \ a_{\textbf{k}'}]=0, \ \ \
 [a_{\textbf{k}}, \ a_{\textbf{k}'}^{\dagger}]=
 \delta_{\textbf{k}, \textbf{k}'}; \\
 \label{a.40}
 &&[\tilde{a}_{\textbf{k}}, \ \tilde{a}_{\textbf{k}'}]=0, \ \ \
 [\tilde{a}_{\textbf{k}}, \ \tilde{a}^{\dagger}_{\textbf{k}'}]=
 \delta_{\textbf{k}, \textbf{k}'},
\end{eqnarray}
with all commutators between the tilde and non-tilde operators vanishing.

Just as in the case of fields, we may introduce the
thermal doublet notation for the operators $a_{\textbf{k}}$ and
$\tilde{a}_{\textbf{k}}$ and then write the Green's function
$G^{\alpha\beta}(\textbf{x},t;\textbf{x}',t')$
in the single-particle momentum representation as
\begin{equation}
 \label{a.41}
 G^{\alpha\beta}(\textbf{x},t;\textbf{x}',t')=\frac{1}{(2\pi)^3}
 \int d\textbf{k} \int d\textbf{k}'
 \phi_{\textbf{k}}^{\alpha}(\textbf{x})
 {\phi}_{\textbf{k}'}^{\beta\dagger}(\textbf{x}')
 \langle 0, \beta| T[a_{\textbf{k}}^{\alpha}(t),a_{\textbf{k}'}^{\beta\dagger}(t')]
 |0, \beta\rangle.
\end{equation}
In the case of a uniform system governed by a time-independent Hamiltonian,
the single-particle states are plane waves
\begin{eqnarray}
 \label{a.42}
 \phi_{\textbf{k}}(\textbf{x})&=&\frac{1}{\sqrt{V}}\exp(i \textbf{k}\cdot \textbf{r}), \\
 \label{a.43}
 \tilde{\phi}_{\textbf{k}}(\textbf{x}) &=&
 \frac{1}{\sqrt{V}}\exp(-i \textbf{k} \cdot\textbf{r}).
\end{eqnarray}
Similarly, $G^{\alpha\beta}(\textbf{x},t;\textbf{x}',t')$  now
depends only on the differences $\textbf{x}-\textbf{x}'$
and $t-t'$, so that
\begin{equation}
 \label{a.44}
 G^{\alpha\beta}(\textbf{x},t)=\int d\textbf{k} \
 e^{i\textbf{k}\cdot\textbf{x}} G^{\alpha\beta}(\textbf{k},t),
\end{equation}
where
\begin{equation}
 \label{a.45}
 G^{\alpha\beta}(\textbf{k},t)\equiv\langle 0, \beta|
 T[a_{\textbf{k}}^{\alpha}(t), a_{\textbf{k}}^{\beta\dagger}(0)]
 0, \beta\rangle
\end{equation}
is the single-particle Green's function in the momentum representation.

The remaining task is to calculate the Green's function
$G^{\alpha\beta}(\textbf{k},t)$. For a system of Bose particles,
this task is complicated by the possible phase transition to a Bose-condensed
state by spontaneous symmetry breaking below a certain critical
temperature $T_C$. This is signalled by a non-vanishing average
\begin{equation}
 \label{a.46}
 \langle 0, \beta| \psi^{\alpha}(\textbf{x})|0, \beta\rangle \equiv
 \Phi_0^{\alpha}(\textbf{x})\neq 0,
\end{equation}
The function $\Phi_0^{\alpha}(\textbf{x})$
is often referred to as the macroscopic wave function
of the condensate and in general has an amplitude and phase. In a uniform
system and in the absence of any supercurrent, we can take $\Phi_0^{\alpha}(\textbf{x})$
to be real and independent of position. In this case, $\Phi_0^{\alpha}(\textbf{x})$ is
equal to the square root of the condensate density $n_0$. We are then led to
separate the boson field operator $\psi^{\alpha}(\textbf{x})$ into two parts:
\begin{eqnarray}
 \label{a.47}
 &&\psi^{\alpha}(\textbf{x})\equiv a^{\alpha}_0/V^{1/2}+
 \bar{\psi}^{\alpha}(\textbf{x}), \\
 \label{a.48}
 &&\psi^{\alpha\dagger}(\textbf{x})\equiv
 a^{\alpha\dagger}_0/V^{1/2}+\bar{\psi}^{\alpha\dagger}(\textbf{x}).
\end{eqnarray}
The commutator of $a_0^{\alpha}$ and $a_0^{\alpha\dagger}$ is unity,
while their product is of order $N$.
Therefore, we may neglect the operator nature of $a_0^{\alpha}$ and
$a_0^{\alpha\dagger}$ and replace them by the $c$-number $\sqrt{N_0}$.
This procedure, known as the Bogoliubov prescription, is appropriate when
the number of particles in the zero-momentum state is a finite fraction of $N$.
The error thus introduced is of the order $O(V^{-1})$ and vanishes in the
thermodynamic limit. The new field operators $\bar{\psi}^{\alpha}(\textbf{x}),
\bar{\psi}^{\alpha\dagger}(\textbf{x})$ describe the normal phase atoms,
satisfy Bose commutation relations, and have vanishing vacuum averages:
\begin{equation}
\label{a.49}
 \langle 0, \beta|\bar{\psi}^{\alpha}(\textbf{x})|0, \beta\rangle=
 \langle 0, \beta| \bar{\psi}^{\alpha\dagger}(\textbf{x})|0, \beta\rangle=0.
\end{equation}
With this, the thermal Green's function now becomes:
\begin{equation}
 \label{a.50}
 G^{\alpha\beta}(\textbf{x},t; \textbf{x}', t')=-n_0+
 \bar{G}^{\alpha\beta}(\textbf{x},t; \textbf{x}', t'),
\end{equation}
where $\bar{G}^{\alpha\beta}(\textbf{x},t; \textbf{x}', t')$ is
its non-condensate part defined by
\begin{equation}
 \label{a.51}
 \bar{G}^{\alpha\beta}(\textbf{x},t; \textbf{x}', t')\equiv -
 \langle 0, \beta|T [\bar{\psi}^{\alpha}(\textbf{x},t),
 \bar{\psi}^{\beta\dagger}(\textbf{x}',t')]|0, \beta\rangle.
\end{equation}
In the momentum representation for $\textbf{k}\neq 0$, we have
\begin{equation}
 \label{a.52}
 \bar{G}^{\alpha\beta}(\textbf{k},t)=
 -\langle 0, \beta| T [a_{\textbf{k}}^{\alpha}(t),
 a_{\textbf{k}}^{\beta\dagger}(0)]|0, \beta\rangle.
\end{equation}
The Bogoliubov prescription modifies the Hamiltonian (\ref{a.27}) in a fundamental way.
In order to simplify matters, we will concentrate on the
zero-temperature part $K$ for the moment, making a straightforward
generalization to TFD later on.
After all of the above transformations, $K$ reads
\begin{equation}
 \label{a.53}
 K=E_0-\mu N_0+\frac{1}{V}\sum_{\textbf{k}\neq 0}(\varepsilon_{\textbf{k}}
-\mu)a_{\textbf{k}}^{\dagger}a_{\textbf{k}}+
\sum_{j=1}^{7}\hat{V}_j,
\end{equation}
where the interaction term $\hat{V}$ can be separated into the following
eight distinct parts:
\begin{eqnarray}
 \label{a.54}
&& E_0=\frac{1}{2}n_0^2 V U(0), \\
 \label{a.55}
 && \hat{V}_1=\frac{1}{2}n_0\sum_{\textbf{k}}
 U(\textbf{k})a_{\textbf{k}}a_{-\textbf{k}}, \\
 \label{a.56}
 && \hat{V}_2=\frac{1}{2}n_0\sum_{\textbf{k}}
 U(\textbf{k})a^{\dagger}_{\textbf{k}}a_{-\textbf{k}}, \\
 \label{a.57}
&& \hat{V}_3=n_0\sum_{\textbf{k}}
 U(\textbf{k})a_{\textbf{k}}^{\dagger}a_{\textbf{k}}, \\
 \label{a.58}
&& \hat{V}_4=n_0\sum_{\textbf{k}}
 U(\textbf{0})a_{\textbf{k}}^{\dagger}a_{\textbf{k}}, \\
 \label{a.59}
&& \hat{V}_5=\frac{n_0^{1/2}}{V^{1/2}}\sum_{\textbf{k},\textbf{q}}
 U(\textbf{q})a_{\textbf{k}+\textbf{q}}a_{\textbf{k}}a_{\textbf{q}}, \\
 \label{a.60}
&& \hat{V}_6=\frac{n_0^{1/2}}{V^{1/2}}\sum_{\textbf{k},\textbf{q}}
 U(\textbf{q})a_{\textbf{k}}^{\dagger}
 a_{\textbf{q}}^{\dagger}a_{\textbf{k}+\textbf{q}}, \\
 \label{a.61}
&& \hat{V}_7=\frac{1}{2 V}\sum_{\textbf{k},\textbf{k}',\textbf{q}}
 U(\textbf{q})a_{\textbf{k}+\textbf{q}}^{\dagger}
 a_{\textbf{k}'-\textbf{q}}^{\dagger}
 a_{\textbf{k}'}a_{\textbf{k}}
\end{eqnarray}
with
\begin{equation}
 \label{a.62}
 U(\textbf{q})\equiv\int d\textbf{x} \ e^{i\textbf{q}\cdot\textbf{x}}
 U(\textbf{x}).
\end{equation}
In a normal system ($n_0=0$), only $\hat{V}_7$ is present. We also note that
$K$ has no term containing a single $a_{\textbf{k}}^{\dagger}$
or $a_{\textbf{k}}$ because these would violate momentum conservation.
This is in agreement with (\ref{a.49}), which gives in a momentum representation
\begin{equation}
 \label{a.63}
 \langle 0, \beta| a_{\textbf{k}} |0, \beta \rangle=
 \langle 0, \beta| a_{\textbf{k}}^{\dagger} |0, \beta \rangle =0, \ \ \
 (\textbf{k}\neq 0).
\end{equation}

\begin{figure}[H]
\begin{center}
\includegraphics[width=10cm, angle=0]{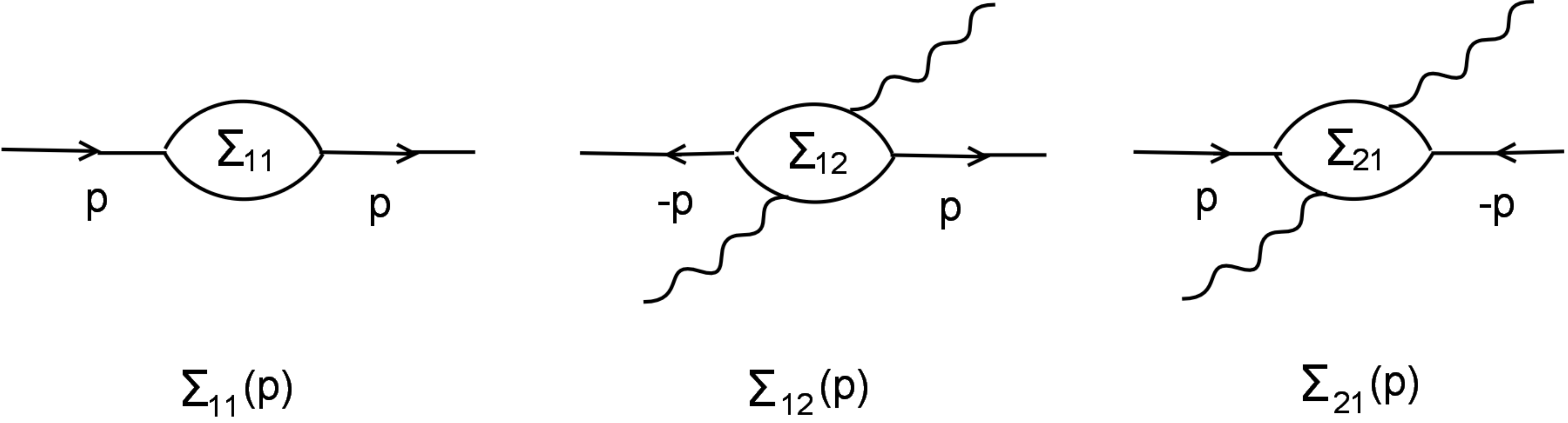}
\end{center}
\caption{Proper self-energies for a Bose-condensed system. Straight lines
 correspond to the normal phase particles, with wavy lines associated
 with the condensed phase.}
\label{fig:Fig.1.10.1}
\end{figure}

\begin{figure}[H]
\begin{center}
\includegraphics[width=8cm, angle=0]{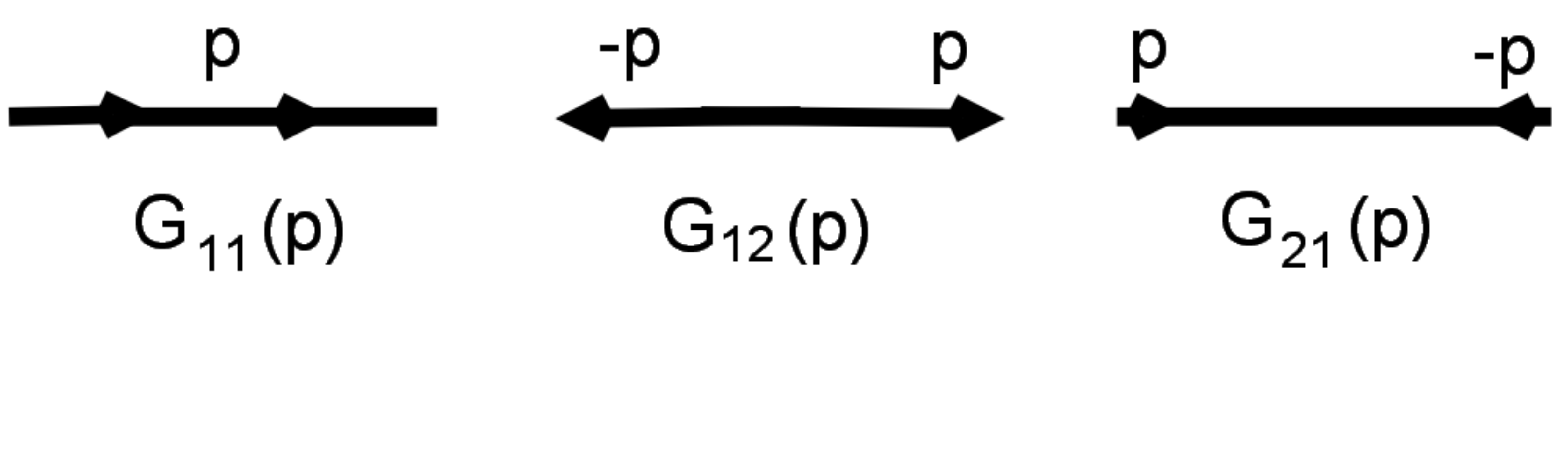}
\end{center}
\caption{Green's functions for a Bose-condensed system.}
\label{fig:Fig.1.10.2}
\end{figure}

\begin{figure}[H]
\begin{center}
\includegraphics[width=10cm, angle=0]{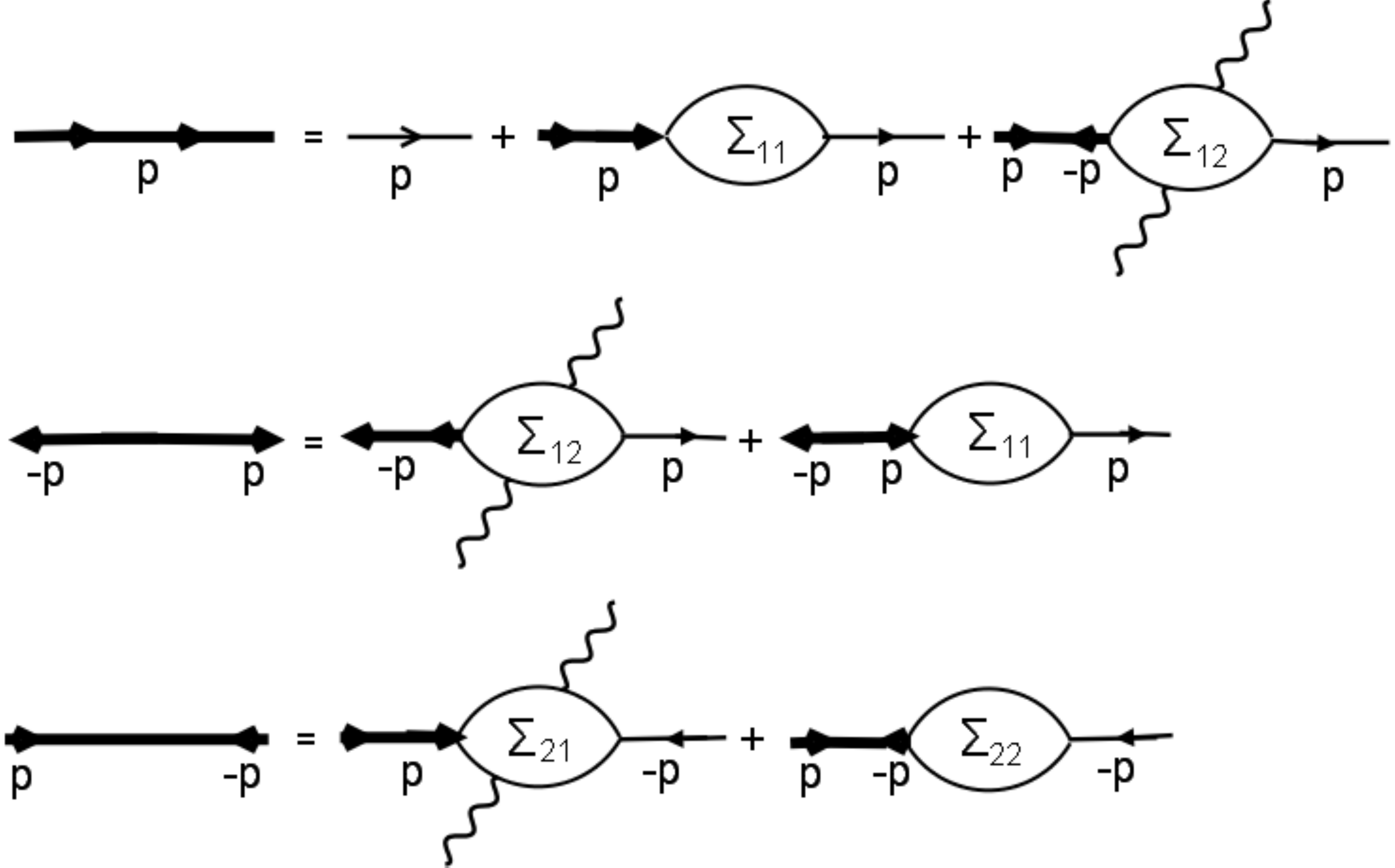}
\end{center}
\caption{The Dyson-Beliaev equations for a Bose-condensed system.
 Thin and thick straight lines denote, respectively, the free and
full propagators of the normal phase particles.}
\label{fig:Fig.1.10.3}
\end{figure}

The fact that there exist interaction terms such as $\hat{V}_1$ implies
the non-conservation of particles in the system described by non--condensate
operators $a_{\textbf{k}}$ and $a_{\textbf{k}}^{\dagger}$ due
to exchanges between non-condensate atoms and the condensate atoms.
This changes the way the proper self-energies are defined. There are three
distinct types of them now, indicated in Fig.~\ref{fig:Fig.1.10.1}.
For $\Sigma_{11}$ there is one particle line coming in and one coming out. The
others components have two particle lines either
coming out $(\Sigma_{12})$ or going in $(\Sigma_{21})$, and reflect the new
features associated with the existence of a Bose condensate.
We then introduce two new Green's functions:
\begin{eqnarray}
 \label{a.64}
 && G_{12}(\textbf{k},t)\equiv -\langle 0, \beta| T a_{-\textbf{k}}(t)
 a_{\textbf{k}}(0) |0, \beta\rangle, \\
 \label{a.65}
&& G_{21}(\textbf{k},t)\equiv -\langle 0, \beta| T a_{\textbf{k}}^{\dagger}(t)
 a_{-\textbf{k}}^{\dagger}(0) |0, \beta\rangle.
\end{eqnarray}
The functions $G_{12}$ and $G_{21}$ are usually called ``anomalous'' Green's functions,
representing the disappearance and appearance of two non-condensate particles,
respectively. The normal Green's function is denoted as $G_{11}$ and
represents the propagation of a single particle
\begin{equation}
 \label{a.66}
 G_{11}(\textbf{k},t)\equiv -\langle 0, \beta| T a_{\textbf{k}}(t)
 a_{\textbf{k}}^{\dagger}(0) |0, \beta\rangle.
\end{equation}
These three Green's functions are shown in Fig.~\ref{fig:Fig.1.10.2}, where the arrows
indicate the direction of momentum of the atoms involved.
The Dyson equations connecting the above free energies and propagators
is found in standard textbooks.
For the case of zero-temperature \cite{Beliaev}, they where first written down by Beliaev,
and are illustrated diagrammatically in Fig.~\ref{fig:Fig.1.10.3}.
The generalization of this to TFD is straightforward and in the momentum representation reads:
\begin{eqnarray}
 \label{a.67}
 &&G_{11}^{\alpha\beta}(p)=G^{(0)\alpha\beta}(p)+
 G_{11}^{\alpha\gamma}(p)\Sigma_{11}^{\gamma\delta}(p)G^{(0)\delta\beta}(p)+
 G_{21}^{\alpha\gamma}(p)\Sigma_{12}^{\gamma\delta}(p)G^{(0)\delta\beta}(p), \\
 \label{a.68}
 &&G_{12}^{\alpha\beta}(p)=
 G_{11}^{\alpha\gamma}(-p)\Sigma_{12}^{\gamma\delta}(p)G^{(0)\delta\beta}(p)+
 G_{12}^{\alpha\gamma}(p)\Sigma_{11}^{\gamma\delta}(p)G^{(0)\delta\beta}(p), \\
 \label{a.69}
 &&G_{21}^{\alpha\beta}(p)=
 G_{11}^{\alpha\gamma}(p)\Sigma_{21}^{\gamma\delta}(p)G^{(0)\delta\beta}(-p)+
 G_{21}^{\alpha\gamma}(p)\Sigma_{22}^{\gamma\delta}(-p)G^{(0)\delta\beta}(p).
\end{eqnarray}
Here the TFD indices $\alpha,\beta$ take the values of 1 and 2 and summation over
repeated indices is implied. The notation $p=(p_0, \textbf{p})$
is employed and the expression for the unperturbed Green's function $G^{(0)\alpha\beta}(p)$
 is given by (\ref{a.24}).
There are some useful relations between the anomalous proper self-energies
and propagators:
\begin{equation}
 \label{a.71}
 G_{12}^{\alpha\beta}(p)=G_{21}^{\alpha\beta}(-p), ~~~
 G_{22}^{\alpha\beta}(p)=G_{11}^{\alpha\beta}(-p), ~~~ 
 \Sigma_{12}^{\alpha\beta}(p)=\Sigma_{21}^{\alpha\beta}(-p), ~~~
  \Sigma_{22}^{\alpha\beta}(p)=\Sigma_{11}^{\alpha\beta}(-p)
\end{equation}
Eqs.~(\ref{a.67}-\ref{a.69}) could be solved with
respect to the propagators in terms of the exact proper self-energies and
unperturbed Green's functions. Therefore, they
are completely general and can be applied to any uniform Bose-condensed fluid,
liquid or gas, without any reference to whether the underlying system is
weakly interacting or not. However, except in special cases, a closed--form
solution is generally not possible, which leads to consideration of various
approximation schemes. Many such schemes, including ours, use
Eqs.~(\ref{a.67}-\ref{a.69})  as a starting point.

\acknowledgments
This work was supported by the Natural Sciences and Engineering Research
Council of Canada.

\end{document}